\documentstyle[multicol,aps,epsfig]{revtex} 

\tighten       

\begin{document}


\title{Quantum transport: The link between standard approaches in 
superlattices}
\author{Andreas Wacker\cite{byline} and Antti-Pekka Jauho}
\address{Mikroelektronik Centret, Bldg 345 east,
Danmarks Tekniske Universitet, 2800 Lyngby, Denmark}
\date{Physical Review Letters {\bf 80}, 369 (1998)}
\maketitle

\begin{abstract}
Theories describing electrical transport in semiconductor superlattices
can essentially be divided in three disjoint categories: i) transport
in a miniband; ii) hopping between Wannier-Stark ladders; and
iii) sequential tunneling.  
We present a quantum transport model,
based on nonequilibrium Green functions, which, in the appropriate limits,
reproduces the three conventional theories, and 
describes the transport  in the  previously unaccessible region
of the parameter space.
\end{abstract}

\begin{multicols}{2}
\narrowtext

Ever since the pioneering work of Esaki and Tsu\cite{ESA70},
which drew attention to the rich physics and
potential device applications of semiconductor
superlattices, these man-made structures have remained
a topic of intense research.
Semiconductor superlattices have proven to be a
fruitful platform for studying a wide range of transport
phenomena, such as their intrinsic negative differential 
conductivity\cite{SIB90},  
the formation of electric field domains \cite{GRA91}, 
Bloch oscillations \cite{WAS93}, as well as dynamical 
localization \cite{HOL92} and
absolute negative conductance\cite{KEA95b} under external irradiation, 
just to mention a few.

\begin{figure}
\noindent\epsfig{file=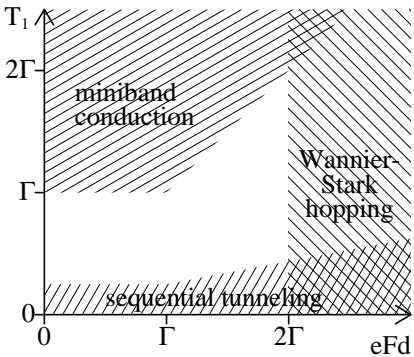,width=5.5cm}\\[0.2cm]
\caption[a]{Regimes where the different transport models
are valid for low electron densities and low temperatures.
(For illustrative purpose we have translated the condition 
$a\gg b$ in the text to $a>2b$, where $a$ and $b$ denote
the respective energy scales involved.)
\label{Figregimes}}
\end{figure}

These phenomena depend crucially on the relations of the
energy scales involved,
namely the zero-field miniband width (which is four times 
the interwell coupling $T_1$), the scattering rate $\Gamma/\hbar$,
and the potential drop per period ($\equiv eFd$, where $F$ is the 
applied static field and $d$ is the superlattice period).
Three distinct approaches
have been used to describe transport in the parameter
space spanned by $(T_1,eFd,\Gamma)$: 
miniband conduction (MBC)\cite{ESA70,LEB70},
Wannier-Stark hopping (WSH)\cite{TSU75}, and sequential 
tunneling (ST)\cite{MIL94,WAC97b}.  
While the ranges of validity of the different approaches have
been addressed qualitatively before \cite{SHI75,TSU91,LAI93,WAC97},
no explicit calculations have been presented where the different ranges
can be identified and the transition between them can be studied.
In the present Letter we present such a calculation, based
on nonequilibrium Green functions. The calculated current-field
relations are shown to reflect the results from the simple
approaches (MBC, WSH, and ST, which will be reviewed below)
in their respective ranges of validity sketched in
the ``phase-diagram'' presented in Fig.~\ref{Figregimes}.
While similar diagrams have been obtained in Refs.~\cite{LAI93,WAC97}
from more phenomenological arguments, we will derive the
borderlines from our Green function analysis here.

Now, we introduce the model assumptions which will
be used in each of the following approaches.
We restrict ourselves to the lowest miniband of the superlattice.
Our basis  set are orthonormal wave-functions 
$\Psi_n(z)e^{i({\bf k}\cdot {\bf r})}/A$ where the $z$-direction denotes 
the growth direction. The $\Psi_n(z)=\Psi(z-nd)$ are localized in 
well $n$ (for example one may use the Wannier-functions). 
Here {\bf r} and {\bf k} denote two-dimensional
vectors within the $(x,y)$-plane (with area $A$)  which is 
assumed to be separable from the $z$-direction.
For  parabolic dispersion $E_k=\hbar^2 {\bf k}^2/2m$
(with the effective mass $m$ of the conduction band) we
thus have a constant density of states 
$\rho_0=m/\pi\hbar^2$ per area and period. 
The single-particle part of the Hamiltonian within nearest neighbor 
coupling is then given by
\begin{equation}
\hat{H}^{\rm SL}_{n,m}=\left(\delta_{n,m-1}+\delta_{n,m+1}\right)T_1
+\delta_{n,m}(E_k-neFd)\label{Eqhamz} \, .
\end{equation}
Furthermore we consider a phenomenological scattering process at 
$\delta$-potentials with density $N_d$ and matrix element
$\delta_{n,m} V/A$,
leading to a scattering rate $1/\tau_0=\Gamma_0/\hbar=N_d\pi V^2\rho_0/\hbar$
between the {\bf k}-states within a given well.  
Finally, we assume that the in-scattering term is 
determined by a Fermi-distribution 
$n_F(E)=[1+\exp((E-\mu)/k_BT_e)]^{-1}$ with electron temperature
$T_e$ and chemical potential $\mu$. 
This assumption establishes internal energy relaxation without
specifying the corresponding processes in detail. It has been has been
implicitly used in the standard approaches\cite{LEB70,TSU75,WAC97b}
as well.

{\em Miniband conduction (MBC):}
For zero electric field Eq.~(\ref{Eqhamz}) is diagonalized by
a set of Bloch functions $\varphi_q(z)=\sum_n e^{inqd}\Psi_n(z)$
and the dispersion relation is given by the miniband
$E(q)=2T_1\cos(qd)$.
The  stationary Boltzmann equation for the distribution function
$f(q,{\bf k})$ is then
\begin{equation}
\frac{eF}{\hbar}\frac{\partial f(q,{\bf k})}{\partial q}=
\frac{n_F(E(q)+E_k)-f(q,{\bf k})}{\tau(E(q)+E_k)}
\label{Eqboltz}\end{equation}
where  the relaxation-time approximation corresponds to our 
assumption on scattering mentioned above. 
For our scattering model, we obtain the relaxation time
$\tau(E)=\tau_0$ for $E\ge2|T_1|$ and 
$\tau(E)=\pi\tau_0/\arccos(-E/2|T_1|)$ for $-2|T_1|\le E<2|T_1|$.
Eq.~(\ref{Eqboltz}) is solved numerically and the current
is calculated from
\begin{equation}
J(F)=\frac{e}{4\pi^3\hbar}\int {\rm d}^2 k\int_{-\pi/d}^{\pi/d}{\rm d}q
f(q,{\bf k})\frac{{\rm d}E(q)}{{\rm d}q}\, .
\end{equation}
The electron density per period is given by
\begin{equation}
N_{2D}=\frac{d}{4\pi^3}\int {\rm d}^2 k\int_{-\pi/d}^{\pi/d}{\rm d}q
f(q,{\bf k})
\end{equation}
and is used to determine the chemical potential (which is field dependent
due to the energy dependence of $\tau(E)$) for a given
electron density. 
This approach can be extended beyond the relaxation time
approximation\cite{IGN91,LEI91}, but the generic features
remain unchanged.

{\em Wannier-Stark hopping (WSH):}
In the presence of an
electric field, the eigenstates of the Hamiltonian
become the localized Wannier-Stark states,
\begin{equation}
\phi_{\nu}(z)=\sum_n J_{n-\nu}\left(\frac{2T_1}{eFd}\right)
\Psi_n(z)\label{EqWSstate}
\end{equation}
with energy $E_{\nu}=-\nu eFd$ where $J_n(z)$ is the Bessel function of the
first kind. Scattering causes hopping between the different states.
Within Fermi's golden rule, the current is given by
\begin{eqnarray}
J(F)&=&\sum_{l>0} l\frac{e}{\tau_0}  
\sum_n\left[J_{n}\left(\frac{2T_1}{eFd}\right)
J_{n-l}\left(\frac{2T_1}{eFd}\right)\right]^2 \nonumber \\
&&\times \frac{1}{2\pi^2}\int{\rm d}^2 k \left[n_F(E_k)-n_F(E_k+leFd)\right]\, .
\end{eqnarray}
Here the term $\sum [J_{n}J_{n-l}]^2$ arises due to the spatial overlap 
of the Wannier-Stark functions  and the Fermi-functions
reflect our assumption regarding in-scattering.
The electron density per period is given by:
\begin{equation}
N_{2D}=\rho_0k_BT_e\log\left[1+\exp\left(\frac{\mu}{k_BT_e}\right)\right]
\end{equation}
which relates $\mu$ to $N_{2D}$.
Again, it is possible to generalize this approach to
more realistic scattering 
mechanisms\cite{ROT97,BRY97}.

{\em Sequential tunneling (ST):}
In this approximation the phase information is
lost after each tunneling event between adjacent wells.
The scattering within a well
is treated self-consistently by 
solving for the spectral functions
$A({\cal E},{\bf k})$; in this work we use the
self-consistent Born-approximation\cite{MAH90} 
for the self-energy.
The transitions to neighboring wells are
calculated in lowest order of the coupling yielding\cite{WAC97,MAH90,MUR95}:
\begin{eqnarray}
J(F)&=&\frac{e}{2\pi^2}\int{\rm d}^2 k 
\int\frac{{\rm d}{\cal E}}{2\pi\hbar } T_1^2 A({\cal E},{\bf k})
A({\cal E}+eFd,{\bf k})\nonumber\\
&& \times \left[n_F({\cal E})-n_F({\cal E}+eFd)\right]\;.
\end{eqnarray}
The carrier density is given by:
\begin{equation}
N_{2D}=\frac{1}{2\pi^2}\int{\rm d}^2 k
\int\frac{{\rm d}{\cal E}}{2\pi} n_F({\cal E})A({\cal E},{\bf k})\, .
\end{equation}
This approach gives quantitative agreement with experiments
in weakly coupled structures
when realistic models for impurity and interface
scattering are employed\cite{WAC97b,WAC97}.

The important issue to recognize is that these three approaches 
treat scattering, external field, and coupling within
different approximations. 
MBC does not properly include field-induced 
localization because of its inherent 
assumption of extended states, WSH treats scattering in lowest order 
perturbation theory (in particular, there is no 
broadening of the states), and ST is explicitly lowest order in 
the interwell coupling.
In contrast to these shortcomings, a full quantum transport theory,
based on nonequilibrium Green functions\cite{HAU96}, 
is able to treat scattering, electric field, and coupling on 
equal footing. Such an approach was performed in Ref.~\cite{LAI93}
using a basis of Wannier-Stark states and restricting the analysis  
to a high  electron temperature.
Here we work within the basis $\Psi_n(z)$ and consider the general 
situation which allows an analysis of transitions between the 
simplified approaches MBC, WSH, and ST. 

{\em Nonequilibrium Green functions (NGF):} Here
the current and electron density are
given by \cite{MAH90,HAU96}
\begin{eqnarray} 
J(F)&=&\frac{e}{2\pi^2}\int{\rm d}^2 k \frac{2}{\hbar}\,{\rm Re}
\left\{T_1 G^<_{n+1,n}(t,t,{\bf k})\right\}\label{Eqneqstrom}\\
N_{2D}&=&\frac{1}{2\pi^2}\int{\rm d}^2 k 
G^<_{n,n}(t,t,{\bf k})\label{Eqneqdichte}
\end{eqnarray}
where  $G^<_{m,n}(t,t',{\bf k})=
i\langle a_{n}^{\dag}(t',{\bf k})a_{m}(t,{\bf k})\rangle$,
and $a_n^{\dag}(t,{\bf k})$ and $a_n(t,{\bf k})$ are the creation
and annihilation operators for the state
$\Psi_n(z)e^{i({\bf k}\cdot {\bf r})}/A$ in well $n$. 
We also need the retarded Green function:
$G^{\rm ret}_{m,n}(t,t',{\bf k})=-i\Theta(t-t')
\langle \{a_{m}(t,{\bf k}),a_{n}^{\dag}(t',{\bf k})\}\rangle$,
where $\{A,B\}$ denotes the anticommutator.    
In the stationary state the Green functions only depend
on the time difference $\tau=t-t'$, and we define the Fourier
transformation via\cite{HAU96}:
\begin{equation}
G_{m,n}({\cal E},{\bf k})=\int {\rm d}\tau 
e^{i\left({\cal E}-eFd\frac{n+m}{2}\right)\frac{\tau}{\hbar}}
G_{m,n}(t,t-\tau,{\bf k})\, .
\end{equation}
Without scattering between the {\bf k}-states and at $T_1=0$
the Green-functions are diagonal in the well index: 
$G^{\rm ret}_{m,n}({\cal E},{\bf k})=
\delta_{m,n}g^{\rm ret}_{n}({\cal E},{\bf k})$
with the free particle Green-function 
$g^{\rm ret}_{n}({\cal E},{\bf k})=1/({\cal E}-E_k+i0^+)$.
The full Green function is then determined by the Dyson equation:
\begin{eqnarray}
G_{m,n}^{\rm ret}({\cal E},{\bf k})&=&
g^{\rm ret}_{m}\left({\cal E}+eFd\frac{m-n}{2},{\bf k}\right)\nonumber \\
&&\big[\delta_{m,n} + \sum_l\Sigma_{m,l}^{\rm ret}
\left({\cal E}+eFd\frac{l-n}{2},{\bf k}\right)\nonumber \\
&&\times G_{l,n}^{\rm ret}\left({\cal E}+eFd\frac{l-m}{2},{\bf k}\right)\big]
\, .
\label{Eqdyson} 
\end{eqnarray}
Within the self-consistent Born approximation for the
scattering the self-energy can be written as
\begin{equation}
\Sigma_{m,n}^{\rm ret}({\cal E},{\bf k})=
\delta_{m,n}\tilde{\Sigma}^{\rm ret}_n({\cal E},{\bf k})
+T_1\delta_{m+1,n}+T_1\delta_{m-1,n}\label{Eqsigma}
\end{equation}
with
$\tilde{\Sigma}^{\rm ret}_n({\cal E},{\bf k})=
N_d/A\sum_{{\bf k}'}V^2G_{n,n}^{\rm ret}({\cal E},{\bf k}')$.
If the scattering term $\tilde{\Sigma}^{\rm ret}_n$ is 
neglected, the solution corresponds to the Wannier-Stark states 
(\ref{EqWSstate}).
On the other hand, neglecting the coupling $T_1$ gives the
spectral functions used in the sequential tunneling model.
Eqs.~(\ref{Eqdyson},\ref{Eqsigma}) are solved self-consistently
for $G^{\rm ret}$. Then $G^{<}$ is calculated via the Keldysh
equation\cite{HAU96}:
\begin{eqnarray}
G^{<}_{m,n}({\cal E},{\bf k})&=&\sum_{m_1}
G^{\rm ret}_{m,m_1}\left({\cal E}+eFd\frac{m_1-n}{2},{\bf k}\right)\nonumber \\ 
&&\phantom{\sum_{m_1}}\times \tilde{\Sigma}^{<}_{m_1}
\left({\cal E}+eFd\left(m_1-\frac{m+n}{2}\right),{\bf k} 
\right)\nonumber \\ 
&&\phantom{\sum_{m_1}}\times G^{\rm adv}_{m_1,n}
\left({\cal E}+eFd\frac{m_1-m}{2},{\bf k}\right) \, .
\label{EqGlessom}
\end{eqnarray}
According to our general assumption about in-scattering we replace
$\tilde{\Sigma}^{<}_{m}({\cal E},{\bf k})$  by its equilibrium value
$-2in_F({\cal E}) 
{\rm Im}\{\tilde\Sigma_{m}^{\rm ret}({\cal E},{\bf k})\}$.
Finally, the current and electron density are calculated via 
Eqs.~(\ref{Eqneqstrom},\ref{Eqneqdichte}).
The extension of this model  to more realistic scattering
processes is straightforward by using the respective self-energies in 
Eqs.~(\ref{Eqsigma},\ref{EqGlessom}) and relaxing
the assumption about in-scattering, although
the calculations become very tedious (see, e.g., Ref.~\cite{LAK97}
where NGF has been applied to the resonant tunneling diode).

\begin{figure}
\noindent\epsfig{file=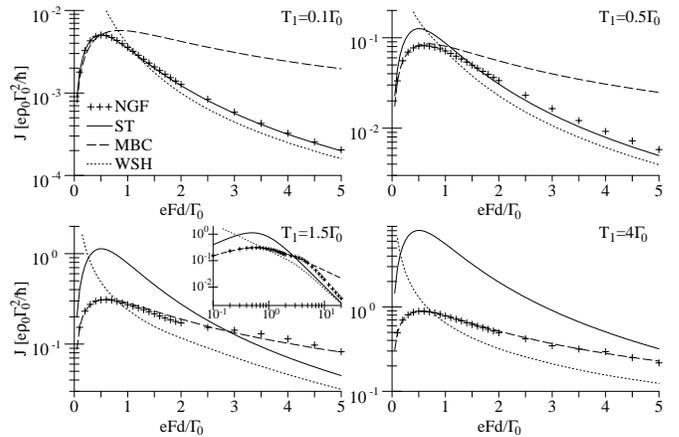,width=8.7cm}\\[0.2cm]
\caption[a]{Current-field relations calculated from nonequilibrium 
Green functions (NGF) in comparison with the standard approaches
for $N_{2D}=0.2\Gamma_0 \rho_0$ and $k_BT_e=0.2\Gamma_0$. 
For $T_1=1.5\Gamma_0$  the current-field relation is also
shown over a wider field range  in the inset.
Here one can see explicitly,
that the NGF result leaves the MBC curve for $eFd\gtrsim T_1$
and approaches the ST and WSH curves for large fields.
\label{FiglowTEF}}
\end{figure}

{\em Results:}
In Fig.~\ref{FiglowTEF} we display the evolution of the current-field
relations for the different models from weakly to strongly-coupled
superlattices.
The curves for MBC, ST, and NGF are qualitatively
similar for all couplings. For low electric fields the current increases
linearly with the electric field. Then there is a peak at intermediate
fields and negative differential conductivity occurs at higher field. 
For small $T_1$  as well as for high fields
the result from ST is in quantitative agreement with
the NGF result, while the results deviate for larger $T_1$.
In contrast, the result from MBC is in quantitative agreement
with the NGF result for large $T_1$ and small $eFd$. 
The WSH-result diverges for $eFd\to 0$ \cite{ROT97},
but approaches the NGF result for large fields. 
These results as well as further
calculations for $k_BT_e<\Gamma$,$N_{2D}<\rho_0\Gamma$
are summarized by Fig.~\ref{Figregimes}, depicting
the respective regions in parameter space, 
where the different approaches approximate the NGF result.

Now we want to justify these ranges of validity
by studying the quantum mechanical correlation between  the wells
$n$ and $m$ given by the retarded Green function. 
For a constant scattering self-energy 
$\tilde{\Sigma}_{n}^{\rm ret}({\cal E},{\bf k})=-i\Gamma/2$
we have found an analytic solution
of Eqs.~(\ref{Eqdyson},\ref{Eqsigma})
\begin{equation}
G^{\rm ret}_{m,n}({\cal E},{\bf k})=\sum_{\alpha}
\frac{J_{m-\alpha}\left(\frac{2T_1}{eFd}\right)
J_{n-\alpha}\left(\frac{2T_1}{eFd}\right)}
{{\cal E}-eFd\left(\frac{m+n}{2}-\alpha\right)-E_k+i\frac{ \Gamma}{2}}
\label{EqGretgamma}
\end{equation}
which is a superposition of broadened Wannier-Stark states.
The Wannier-Stark ladder  becomes resolved if 
$eFd\gg  \Gamma$. This  defines the
region of validity for the WSH-approach, 
as indicated by the right region in Fig.~\ref{Figregimes}.
By Fourier transforming we obtain
\begin{eqnarray}
G^{\rm ret}_{m,n}(t,t-\tau,{\bf k})&=&-i \Theta(\tau)i^{n-m}
e^{i\left(\frac{m+n}{2}eFd-E_k\right)\frac{\tau}{\hbar}}
e^{- \Gamma \tau/2\hbar}\nonumber \\
&&\times J_{m-n}\left[\frac{4T_1}{eFd}\sin 
\left(\frac{eFd}{2\hbar}\tau\right)\right] \, . \label{EqGretgammat}
\end{eqnarray}
Here the terms $G^{\rm ret}_{n\pm 1,n}$ become of the order
of $G^{\rm ret}_{n,n}$ when $|J_{0}|\approx |J_{\pm 1}|$, i.e., 
$\left|4T_1/(eFd)\sin \left(eFd\tau/2\right)\right|
\approx \sqrt{2}$ . This can be used as an estimate for  the boundary
between localization and delocalization. 
Due to the exponential factor in Eq.~(\ref{EqGretgammat})
only $\tau<2\hbar/ \Gamma$ is of relevance. If $eFd> \Gamma$ the magnitude
of the sine takes the
average value $\approx 1/\sqrt{2}$. Then we find  $2|T_1|\approx eFd$.
If, on the other hand, $eFd< \Gamma$ we may replace  $\sin(x)\approx x$ and 
have  $2|T_1|\approx  \Gamma$
at the time $\tau=\sqrt{2}\hbar/\Gamma$. From these estimates 
we conclude that the states are essentially delocalized if
$2|T_1|\gg  \Gamma$ and $2|T_1|\gg eFd$.
In this case the miniband states form a useful basis as indicated
in the upper left part of Fig.~\ref{Figregimes}.
On the other hand for
$2|T_1|\ll  \Gamma $ or $2|T_1|\ll eFd$
the correlation functions $G^{\rm ret}_{m,n}$ vanish for $m\neq n$
and the states are essentially localized so that the sequential limit
can be used,  as indicated
in the lower part of Fig.~\ref{Figregimes}.

\begin{figure}
\noindent\epsfig{file=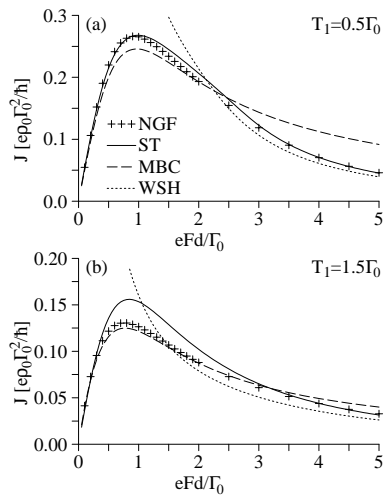,width=5.0cm}\\[0.2cm]
\caption[a]{Current-field relations for (a) high electron density
($N_{2D}=2\Gamma_0\rho_0$ and $k_BT_e=0.2\Gamma_0$) 
and (b) high electron temperature 
($k_BT_e=3\Gamma_0$ and $N_{2D}=0.2\Gamma_0\rho_0$)
for the different
approaches.
Note, that the MBC result deviates from the NGF result at
$eFd\gtrsim N_{2D}/\rho_0$ in (a).
\label{FighighTEF}
}
\end{figure}

For larger electron densities the agreement
between the different approaches becomes better
as shown in Fig.~\ref{FighighTEF}(a).
These results together with further calculations
indicate that ST is also valid
if $N_{2D}/\rho_0\gtrsim 2|T_1|$, and MBC
is also valid if $N_{2D}/\rho_0\gtrsim \Gamma$ and 
$N_{2D}/\rho_0\gtrsim  eFd$.
A similar trend is found for  higher electron temperatures 
(Fig.~\ref{FighighTEF}(b)). This agrees with the analytic findings
of Ref.~\cite{LAI93} where it is shown that NGF gives the same result 
as MBC in the limit $k_BT_e\gg |T_1|,eFd$.

{\em In conclusion} we have explicitly shown
that a transport calculation based on nonequilibrium 
Green functions contains the simple approaches MBC, WSH, and ST 
as limiting cases.
For low temperature and low electron density the ranges of validity
of the simplified approaches are depicted in Fig.~\ref{Figregimes}, 
while for higher electron densities or temperatures, these ranges
are enlarged.
The essential message of our analysis is that for wide regions
in parameter space (but not everywhere!) a simplified theory can
be found, which approximates the full theory satisfactorily.  
This should have important consequences for practical device 
modeling, where other complications,
such as realistic scattering mechanisms, must be considered as well.

A.W. acknowledges financial support by the
Deutsche Forschungsgemeinschaft.

\end{multicols}

\end{document}